\documentclass[12pt,a4paper]{article}
\usepackage{amsmath,epsfig}
\begin{document}
 
\setlength{\unitlength}{.8mm}
\newcommand{\be}{\begin{equation}}
\newcommand{\bea}{\begin{eqnarray}}

\newcommand{\Eps}{\Epsilon}
\newcommand{\gM}{\mathcal{M}}
\newcommand{\dD}{\mathcal{D}}
\newcommand{\gG}{\mathcal{G}}
\newcommand{\pa}{\partial}
\newcommand{\eps}{\epsilon}
\newcommand{\La}{\Lambda}
\newcommand{\De}{\Delta}
\newcommand{\nonu}{\nonumber}
\newcommand{\beq}{\begin{eqnarray}}
\newcommand{\eeq}{\end{eqnarray}}
\newcommand{\ka}{\kappa}
\newcommand{\an}{\ensuremath{\alpha_0}}
\newcommand{\bn}{\ensuremath{\beta_0}}
\newcommand{\dn}{\ensuremath{\delta_0}}
\newcommand{\al}{\alpha}
\newcommand{\bm}{\begin{multline}}
\newcommand{\fm}{\end{multline}}
\newcommand{\de}{\delta}

\begin{titlepage} 
\vspace*{0.5cm}
\begin{center}
{\Large\bf TBA equations for excited states in the Sine-Gordon model}
\end{center}
\vspace{2.5cm}
\begin{center}
{\large J\'anos Balog and \'Arp\'ad Heged\H us}
\end{center}
\bigskip
\begin{center}
Research Institute for Particle and Nuclear Physics,\\
Hungarian Academy of Sciences,\\
H-1525 Budapest 114, P.O.B. 49, Hungary\\ 
\end{center}
\vspace{3.2cm}
\begin{abstract}
We propose TBA integral equations for multiparticle soliton (fermion)
states in the Sine-Gordon (massive Thirring) model. 
This is based on T-system and Y-system
equations, which follow from the Bethe Ansatz solution in the light-cone 
lattice formulation of the model. Even and odd charge sectors are
treated on an equal footing, corresponding to periodic and twisted
boundary conditions, respectively. The analytic properties of the Y-system  
functions are conjectured on the basis of the large volume solution of
the system, which we find explicitly. A simple relation between the TBA
Y-functions and the counting function variable of the alternative
non-linear integral equation (Destri-deVega equation) description
of the model is given. 
\end{abstract}

\end{titlepage}

\section{Introduction}

A better theoretical understanding of finite size (FS) effects
is one of the most important problems in Quantum Field Theory (QFT).
The study of FS effects is a useful method of analysing the structure
of QFT models and it is an indispensable tool in the numerical
simulation of lattice field theories. 

L\"uscher \cite{Luscher}
derived a general formula for the FS corrections to particle masses
in the large volume limit. This formula, which is generally applicable for
any QFT model in any dimension, expresses the FS mass corrections in 
terms of an integral containing the forward scattering amplitude
analytically continued to unphysical (complex) energy. It is most
useful in $1+1$ dimensional integrable models \cite{KM1}, where the
scattering data are available explicitly.

The usefulness of the study of the mass gap in finite volume is 
demonstrated~\cite{LWW} by the introduction of the L\"uscher-Weisz-Wolff 
running coupling that enables the interpolation between the
large volume (non-perturbative) and the small volume (perturbative)
regions in both two-dimensional sigma models and QCD.

An important tool in the study of two-dimensional integrable field
theories is the Thermodynamic Bethe Ansatz (TBA).
This thermodynamical method was initiated by Yang and Yang \cite{YY} and  
allows the calculation of the free energy of the particle system.
The calculation was applied to the XXZ model by Takahashi and
Suzuki~\cite{TS} who derived the TBA integral equations for
the free energy starting from the Bethe Ansatz solution
of the system and using the \lq\lq string hypothesis'' describing
the distribution of Bethe roots. The form of the resulting TBA
equations strongly depends on whether the anisotropy parameter $p_0$ (see
Section 2) is an integer, rational or irrational number.

The TBA equations also determine FS effects
in relativistic (Euclidean) invariant two-dimensional field theory
models where the free energy is related to the ground state energy in finite
volume by a modular transformation interchanging spatial extension
and (inverse) temperature. Zamolodchikov \cite{Zamo90} initiated
the study of TBA equations for two-dimensional integrable models by
pointing out that TBA equations can also be derived starting from the 
(dressed) Bethe Ansatz equations formulated directly in terms of the
(infinite volume) scattering phase shifts of the particles.
In this approach the FS dependence of the ground state energy 
has been studied \cite{TBAlist} in many integrable models, 
mainly those formulated as perturbations of minimal conformal models.

TBA methods have been developed also in lattice statistical physics
\cite{KP}. Actually the basic functional relations, Y-system equations,
TBA integral equations and also other techniques playing important role
in the TBA analysis of continuum models have been originally
introduced here. TBA equations describing the FS energy of excited
states were proposed for some models. Also non-linear integral equations, 
similar to the Destri-deVega equations, appeared here first. 
An important model, the tricritical Ising model perturbed 
by its $\phi_{1,3}$ operator, has been studied in detail~\cite{PCA}.
The TBA equations describing all excited states are worked out
in this example, both for the massive and the massless perturbed models.

The TBA description of excited states is less systematic in continuum
models. The excited state TBA systems first studied \cite{Fendley,Martins}
are not describing particle states, they correspond to ground states
in charged sectors of the model. An interesting suggestion is to
obtain excited state TBA systems by analytically continuing~\cite{Martins}
those corresponding to the ground state energy. TBA equations
for scattering states were suggested for some perturbed field theory models
by the analytic continuation method \cite{DT}. Excited state TBA
equations were also suggested for scattering multi-particle states
for the Sine-Gordon model at its $N=2$ supersymmetric point \cite{Fendley2}.

The BLZ program \cite{BLZ} has been initiated to derive functional
relations, Y-systems, and TBA integral equations, both for the ground state
and for excited states, directly in the continuum. The construction 
is complete for the conformal field theory limit of the models only
although some examples in massive perturbed models are also worked out.
The original construction is based on the $R$-matrix corresponding to 
the quantum deformed loop algebra $U_q(A^{(1)}_1)$ and can be applied to 
minimal models perturbed by their $\phi_{1,3}$ field but it works also 
in the $U_q(A^{(2)}_2)$ case~\cite{BE}, which is relevant to the
$\phi_{1,2}$ and $\phi_{2,1}$ perturbations. In this family the TBA
equations for the first two excited states in the non-unitary perturbed
${\cal M}_{3,5}$ model are worked out explicitly.

An alternative to the TBA equations is provided by the Destri-deVega (DdV)
non-linear integral equations. They were originally suggested~\cite{DdV0}
for the ground state of the Sine-Gordon (SG) or the closely
related massive Thirring (MT) model, but can be systematically
generalized to describe excited states as well; the (common)
even charge sector of the SG/MT model~\cite{DdV2}, the odd charge
sector in both SG and MT \cite{DdV1} and even the states of perturbed 
minimal models,
which can be represented as restrictions of the SG model \cite{DdVres}. 
The advantage of this non-linear integral equation approach is that
when available (SG/MT model and restrictions), it gives a systematic
description of all excited states. The method has been generalized to
higher rank imaginary coupling affine Toda models~\cite{ZJ}.

Meanwhile an alternative derivation of the TBA equations, 
which is independent of the validity of the string hypothesis 
has been found in the lattice approach for the generalized Hubbard
model \cite{JKS} and also for the XXZ model \cite{Kuniba}. Both models
are solvable on the lattice by the Bethe Ansatz method and the new
derivation of the TBA equations proceeds via the introduction of
T-system and Y-system functions satisfying functional relations
which can be transformed into TBA-type integral equations. Instead of 
the string hypothesis used in the original derivation \cite{TS}, here
the basic assumption is about the analytic structure of the Y-system
functions (distribution of their zeroes) and the validity of these
assumptions can be convincingly demonstrated numerically, at least
for small lattices, by starting from the numerical solution of the 
Bethe Ansatz equations. The Takahashi-Suzuki results for the ground
state energy are reproduced and new TBA equations for the excited states
are found in this way. Actually for the excited state problem the TBA
equations are supplemented by quantization conditions restricting
the allowed values of particle momenta, which is natural for particles
confined in a box.

In this paper we borrow the ideas of the lattice approach and
apply them to the continuum QFT case. Following the steps of the
lattice construction, we systematically build the T-system and Y-system
functions for all multi-soliton (multi-fermion) states of the SG (MT)
model. We find the appropriate continuum TBA equations and
quantization conditions and a link between the TBA and the DdV
equations.  

Our starting point is the light-cone lattice
regularization \cite{LC} of the SG/MT model and the Bethe Ansatz solution.
We use twisted boundary
conditions on the lattice and show that by changing the value of the parameter
characterizing the boundary conditions we can describe both the even
and the odd sectors of the model on an equal footing. Previously only
the even sector was treated in the light-cone lattice formulation
and the properties of the states belonging to the odd sector were
conjectured by postulating the corresponding DdV equation in the
continuum limit \cite{DdV1}. Our result for the odd sector shows that
the conjectured DdV equations are indeed correct in this sector.

Next we introduce the T-system and Y-system functions and 
write down the Y-system functional relations. In the even sector
we recover the usual Y-system equations but we find that in the odd sector
the Y-system equations are somewhat modified. Although the small
modifications (with respect to the even sector) only occur at the
end of the corresponding diagrams, this result shows that a naive
analytic continuation of the even sector TBA equations would lead
to incorrect results for the odd sector.

There is a link between the TBA Y-system functions and the \lq\lq
counting function'' of the DdV approach. 
This link is explicitly given by the simple formula
(\ref{FY}) and is used to write down explicitly the solution
of the TBA Y-system for asymptotically large volume.

To transform the Y-system equations into TBA integral equations (and
quantization conditions) we need to know the analytic properties
of the Y-system functions, in particular the distribution of their zeroes. 
From the explicit solution we know this
distribution in the infinite volume limit.
Our main assumption in this paper is that the qualitative properties
of this distribution remain the same for finite volume. 
In the TBA language the states are characterized by specifying the
source terms in the integral equations and since the functional form of the
source terms
is volume independent, this can be read off from the known solution
in the large volume limit.

Using the above method we write down the complete set of equations
determining multi-particle momenta and energies in the TBA approach.
We have verified the validity of our main assumption by 
numerically comparing the results of the TBA approach with those
obtained using the DdV equations. In a particular case ($p_0=4$),
where we found exact results for one-particle momenta and energies
on the TBA side, we were able to verify the agreement by
numerically solving the corresponding DdV equations to 20 digits.
 
For simplicity, in this paper we restrict attention to the
repulsive case $p_0>2$ of the SG coupling and consider integer $p_0$ only. 
Moreover, we consider multi-soliton states only and no states
containing both solitons and anti-solitons at the same time. We believe
that appropriate TBA systems can be found also for more general
couplings and more general states but the special cases we are
considering in this paper are sufficient to present the main ideas and
assumptions. Finally we note that although in the SG/MT case the
excited state TBA description is \lq\lq superfluous'' since we already
have the DdV equations to study FS physics, one can hope that the
simple pattern of the excited state TBA systems we find here means 
that similar systems can be found also for other models, where no DdV
type alternative is available.

The paper is organized as follows.
The light-cone lattice regularization of the SG/MT model and the Bethe 
Ansatz solution is briefly recalled in Section 2. 
In Section 3 we introduce
the T-system and Y-system functions and establish the link between
the TBA Y-system and the \lq\lq counting function'' of the DdV
approach. In Section 4 simple properties of the counting function
are recalled. In Section 5 we write down the light-cone lattice
TBA equations together with the quantization conditions. In section 6
we take the (finite volume) continuum limit of the equations. 
Multi-particle energy and momentum expressions are given in Section 7.
In Section 8 we discuss a special case ($p_0=4$) where we can find
exact expressions for the one-particle energies and momenta.
In section 9 we discuss the DdV equation and its analytic continuation
to the entire complex rapidity plane. In Section 10, using the link
between the DdV equation and the TBA Y-system, we find the explicit 
solution of the infinite volume limit of the TBA problem. 
The ground state (no particles), one-particle, and two-particle
problems are discussed in detail in Sections 11, 12, and 13, 
respectively. The numerical comparison of the results of the TBA
approach with those obtained using the DdV equations is briefly
described in Section~14 and finally our conclusions are summarized 
in Section 15. The technical details of the transformation of 
Y-system type functional relations into
TBA type integral equations is discussed in the Appendix.


\section{Light-cone approach to twisted SG model}
Our starting point is the Bethe Ansatz solution of the integrable lattice
regularization of the Sine-Gordon field theory \cite{LC}. Here we briefly
summarize the results of this approach \cite{DdV0,DdV2,DdV1} 
to the Sine-Gordon (and
massive Thirring) model. The fields of the regularized theory
are defined at sites (\lq\lq events") of a light-cone lattice and the
dynamics of the system is defined by 
translations in the left and right lightcone directions. These are given by
transfer matrices of the six-vertex model with anisotropy $\gamma$ and
alternating inhomogeneities. This approach is particularly useful for
calculating the finite size dependence of physical quantities.
We take $N$ points ($N$ even) in the spatial direction and use twisted periodic
boundary conditions. The lattice spacing is related to $L$, the
(dimensionful) size of the system:
\be
a=\frac{L}{N}.
\end{equation}

The physical states of the system are characterized by the set of
Bethe roots\break $\{w_j,\ j=1,\dots,m\leq\frac{N}{2}\}$, which satisfy the 
Bethe Ansatz equations (BAE) 
\be
\frac{Q(w_j+2i)}{Q(w_j-2i)}=
-\frac{T_0(w_j+i)}{T_0(w_j-i)}
\qquad\qquad j=1,\dots,m.
\label{BAE}
\end{equation}

Here
\be
T_0(\xi)=\left[
k\left(\xi-\frac{2\Theta}{\pi}\right)
k\left(\xi+\frac{2\Theta}{\pi}\right)\right]^{N/2},
\end{equation}
where $\Theta$ is the inhomogeneity parameter,
\be
k(\xi)=\frac{\sinh\frac{\gamma}{2}\xi}{\sin\gamma}
\end{equation}
and
\be
Q(\xi)=e^{\frac{\omega\xi}{2}}\,\prod_{j=1}^m\,
\sinh\frac{\gamma}{2}(\xi-w_j),
\end{equation}
where the parameter $\omega$ is characterizing the twist of the
boundary condition.
We shall also use the parametrization
\be
\gamma=\frac{\pi}{p_0}=\frac{\pi}{p+1}.
\end{equation}
Finally the energy ($E$) and momentum ($P$) of the physical state can be
obtained from the eigenvalues of the lightcone transfer matrices:
\be
e^{ia(E\pm P)}=(-1)^m\, e^{\pm2i\omega}\,
\frac{Q\left[\pm\left(-i +\frac{2\Theta}{\pi}\right)\right]}
{Q\left[\pm\left(i +\frac{2\Theta}{\pi}\right)\right]}.
\label{EP}
\end{equation}

Besides the usual procedure, taking the thermodynamic limit
($N\to\infty$) first, followed by the continuum limit ($a\to0$)
one can also study continuum limit in finite volume by taking
$N\to\infty$ and tuning the inhomogeneity parameter $\Theta$
simultaneously as
\be
\Theta=\ln\frac{2}{{\cal M}a}=\ln\frac{2N}{l},
\label{Theta}
\end{equation}
where the mass parameter ${\cal M}$ is the
infinite volume physical mass of the Sine-Gordon solitons
and we have introduced $l={\cal M}L$, the dimensionless size of the system.

\section{T-system and Y-system}
\newsavebox{\Asp}
\sbox{\Asp}{\begin{picture}(52,5)(0,-3.5)
\put(10,0){\circle*{3}}
\multiput(20,0)(10,0){4}{\circle{3}}
\put(10,0){\line(1,0){8.5}}
\multiput(22,0)(1,0){7}{\circle*{.2}}
\put(31.6,0){\line(1,0){7}}
\put(43.7,0.4){\line(1,0){4.8}}
\put(43.7,-0.4){\line(1,0){4.8}}
\put(41.5,0){\line(4,1){5}}
\put(41.5,0){\line(4,-1){5}}
\put(12,-2){\makebox(0,0)[t]{{\protect\scriptsize 1}}}
\put(22,-2){\makebox(0,0)[t]{{\protect\scriptsize 2}}}
\put(32,-2){\makebox(0,0)[t]{{\protect\scriptsize {\em p}--2}}}
\put(42,-2){\makebox(0,0)[t]{{\protect\scriptsize {\em p}--1}}}
\put(52,-2){\makebox(0,0)[t]{{\protect\scriptsize {\em p}}}}
\put(58,-2){$b$}

\end{picture}}
\newsavebox{\Dp}
\sbox{\Dp}{\begin{picture}(52,5)(0,-3.5)
\put(10,0){\circle*{3}}
\multiput(20,0)(10,0){4}{\circle{3}}
\multiput(11.5,0)(10,0){2}{\line(1,0){7}}
\multiput(32.5,0)(1,0){6}{\circle*{.2}}
\put(41.5,0){\line(1,0){7}}
\put(51.1,1.1){\line(1,1){7.7}}
\put(51.1,-1.1){\line(1,-1){7.7}}
\put(60,10){\circle{3}}
\put(60,-10){\circle{3}}
\put(65,10){\makebox(0,0)[t]{{\protect\scriptsize {\em p}+1}}}
\put(64,-10){\makebox(0,0)[t]{{\protect\scriptsize {\em p}}}}
\put(12,-2){\makebox(0,0)[t]{{\protect\scriptsize 1}}}
\put(22,-2){\makebox(0,0)[t]{{\protect\scriptsize 2}}}
\put(32,-2){\makebox(0,0)[t]{{\protect\scriptsize 3}}}
\put(41,-2){\makebox(0,0)[t]{{\protect\scriptsize {\em p}--2}}}
\put(48,-2){\makebox(0,0)[t]{{\protect\scriptsize {\em p}--1}}}
\put(67,-2){$a$} 
\end{picture}}
\newsavebox{\Ap}
\sbox{\Ap}{\begin{picture}(52,5)(0,-3.5)
\put(10,0){\circle*{3}}
\multiput(20,0)(10,0){4}{\circle{3}}
\multiput(11.5,0)(10,0){2}{\line(1,0){7}}
\multiput(32.5,0)(1,0){6}{\circle*{.2}}
\put(41.5,0){\line(1,0){7}}
\put(12,-2){\makebox(0,0)[t]{{\protect\scriptsize 1}}}
\put(22,-2){\makebox(0,0)[t]{{\protect\scriptsize 2}}}
\put(32,-2){\makebox(0,0)[t]{{\protect\scriptsize 3}}}
\put(41,-2){\makebox(0,0)[t]{{\protect\scriptsize {\em p}--3}}}
\put(51,-2){\makebox(0,0)[t]{{\protect\scriptsize {\em p}--2}}}
\end{picture}}

Besides $Q(\xi)$ and $T_0(\xi)$ it is useful to introduce \cite{Kuniba}
the (half-)infinite sequence of functions $T_s(\xi)$ for $s=-1,0,1,\dots$
In the six-vertex model these are eigenvalues of the transfer matrices 
corresponding to higher spin representations (in the auxiliary space)
and are obtained by the fusion procedure. Equivalently we can simply
define them as $T_{-1}(\xi)=0$ and
\be
T_s(\xi)=Q(\xi+i+is)Q(\xi-i-is)\sum_{j=0}^s q\left[\xi+i(2j-s)\right]
\label{Ts}
\end{equation}
for $s=0,1,\dots$, where
\be
q(\xi)=\frac{T_0(\xi)}{Q(\xi-i)Q(\xi+i)}.
\label{q}
\end{equation}
It is obvious from the definition that the $T_s$ are periodic
functions,
\be
T_s(\xi+2ip_0)=T_s(\xi).
\end{equation}
It can also be verified that they satisfy the T-system equations \cite{Kuniba}
\be
T_s(\xi+i)T_s(\xi-i)=
T_{s+1}(\xi)T_{s-1}(\xi)+T_0(\xi+i+is)T_0(\xi-i-is),\quad
s=0,1,\dots
\label{Tsystem}
\end{equation}
which can then be used to recursively calculate them starting from $T_1(\xi)$.

The next step is to introduce the functions
\be
Y_s(\xi)=\frac
{T_{s+1}(\xi)T_{s-1}(\xi)}{T_0(\xi+i+is)T_0(\xi-i-is)},
\label{Y}
\end{equation}
or equivalently
\be
1+Y_s(\xi)=\frac
{T_{s}(\xi+i)T_{s}(\xi-i)}{T_0(\xi+i+is)T_0(\xi-i-is)}.
\label{1Y}
\end{equation}
As consequence of the T-system equations (\ref{Tsystem}) they satisfy
the Y-system equations
\be
Y_s(\xi+i)Y_s(\xi-i)=
[1+Y_{s+1}(\xi)][1+Y_{s-1}(\xi)],\quad
s=1,2,\dots
\label{Ysystem}
\end{equation}

Finally we introduce the exponential counting function 
\be
{\cal F}(\xi)=\frac{Q(\xi+2i)}{Q(\xi-2i)}\,
\frac{T_0(\xi-i)}{T_0(\xi+i)},
\label{exponential}
\end{equation}
which can be used to rewrite the Bethe Ansatz equations as
\be
{\cal F}(w_j)=-1\qquad\qquad j=1,\dots,m.
\label{BAE1}
\end{equation}
Combining the above definitions it is easy to see that the Y-functions
can simply be expressed in terms of the exponential counting function
starting with
\be
1+Y_1(\xi)=\left[1+{\cal F}(\xi+i)\right]
\left[1+\frac{1}{{\cal F}(\xi-i)}\right]
\label{FY}
\end{equation}
and then using the Y-system equations (\ref{Ysystem}) recursively.

The above algebraic considerations are supplemented by the observation
that although the $T_s(\xi)$ as defined by (\ref{Ts}) and (\ref{q})
appear to be
trigonometric rational functions but in fact they are trigonometric
polynomials. This is a consequence of the Bethe Ansatz equations
(\ref{BAE}). It is also easy to see that
\be
T_s(\xi)\sim e^{\frac{\gamma}{2}N\vert\xi\vert}
\label{asy}
\end{equation}
asymptotically for $\xi\to\pm\infty$. The Y-functions are
trigonometric rational functions, bounded for large $\vert\xi\vert$.

In the rest of this paper we restrict our attention to the
following special case of the problem.
\begin{itemize}
\item
$p_0=p+1$ integer ($p\geq2$)
\item
$w_j$ all real and different
\item
$\omega=0$ or $\omega=\frac{\pi}{2}$
\end{itemize}
We have chosen these restrictions because while they make the discussion
technically much easier than for the general case they still contain
many physically interesting cases. We will discuss in detail
\begin{itemize}
\item
case ${\cal A}$: $\omega=0$,
\item
case ${\cal B}_1$: $\omega=\frac{\pi}{2}$, \ \ $p_0$ odd,
\item
case ${\cal B}_2$: $\omega=\frac{\pi}{2}$, \ \ $p_0$ even.
\end{itemize}

For the special case $p_0$ integer the T-functions satisfy 
\be
T_{p_0}(\xi)-T_{p_0-2}(\xi)=2(-1)^m\cos(\omega p_0)T_0(\xi+ip_0)
\label{spec}
\end{equation}
and for cases ${\cal A}$ and ${\cal B}_2$ this allows us to define
\be
K(\xi)=(-1)^m\cos(\omega p_0)\frac{T_{p_0-2}(\xi)}{T_0(\xi+ip_0)},
\label{K}
\end{equation}
which, together with $\{Y_s(\xi),\ 
s=1,2,\dots,p_0-2\}$ form a finite Y-system of type $D_{p_0}$, because
\bea
K(\xi+i)K(\xi-i)&=&1+Y_{p_0-2}(\xi),\\
\left[1+K(\xi)\right]^2&=&1+Y_{p_0-1}(\xi).
\end{eqnarray}
For case ${\cal B}_1$ (\ref{spec}) implies the reflection symmetry
\bea
T_{p_0-1-k}(\xi)&=&T_{p_0-1+k}(\xi)\qquad k=0,1,\dots,p_0\label{refl1}\\
Y_{p_0-1-k}(\xi)&=&Y_{p_0-1+k}(\xi)\qquad k=0,1,\dots,p_0-1\label{refl2}
\end{eqnarray}
and this symmetry allows us to truncate the Y-system after
$Y_{p_0-1}$. We call the corresponding finite Y-system of type
$A_{2p_0-3}^s$.

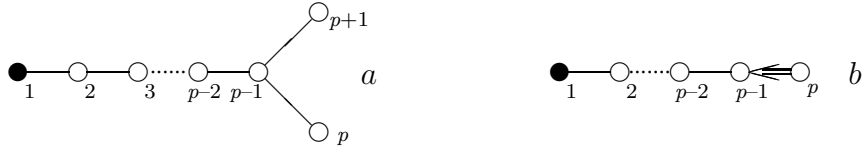
\begin{figure}[htbp]
\begin{center}
\begin{picture}(130,40)(0,-20)
\put(-10,5) {\usebox{\Dp}}
\put(80,5) {\usebox{\Asp}}
\put(-20,-15){\parbox{132mm}{\caption{\label{DpAsp}\protect {\footnotesize
Dynkin-diagrams associated with $D_{p+1}$ and $A^s_{2p-1}$ type 
Y-systems. }}}}
\end{picture}
\end{center}
\end{figure}

To analyse the Y-system equations further we need some definitions.
Let us write
\be
N=2(m+d),
\end{equation}
where $0\leq d\leq\frac{N}{2}$. A special case of ${\cal A}$, for which 
also $d<p_0$ will play an important role in the following. We call
this sub-case $\tilde{\cal A}$. Similarly we denote by
$\tilde{\cal B}_2$ the sub-case of ${\cal B}_2$ with $d<\frac{p_0}{2}$.
We next define $\zeta(\xi)$ by
\be
\zeta(\xi)=\sum_{j=0}^{p_0-1}q\left[\xi+i(2j+1-p_0)\right].
\end{equation}
It is easy to see that for cases ${\cal A}$ and ${\cal B}_2$
\be
\zeta(\xi+2i)=\zeta(\xi)
\end{equation}
and
\be
T_{p_0-1}(\xi)=
\zeta(\xi)Q(\xi+ip_0)Q(\xi-ip_0).
\label{zetaQQ}
\end{equation}
Further
\bea
T_{p_0}(\xi)&=&(-1)^m\kappa T_0(\xi+ip_0)+
\zeta(\xi-i)Q(\xi+i+ip_0)Q(\xi-i-ip_0),\\
T_{p_0-2}(\xi)&=&(-1)^{m+1}\kappa T_0(\xi+ip_0)+
\zeta(\xi-i)Q(\xi+i+ip_0)Q(\xi-i-ip_0),
\end{eqnarray}
where $\kappa=\cos(\omega p_0)$.

\begin{figure}[htbp]
\begin{center}
\begin{picture}(130,30)(0,-20)
\put(35,5) {\usebox{\Ap}}
\put(-20,-7){\parbox{130mm}{\caption{ \label{Ap}\protect {\small
Dynkin-diagram associated with $A_{p-2}$ type Y-systems. }}}}
\end{picture}
\end{center}
\end{figure}
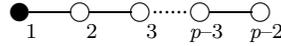
\vspace{-1cm}

Because $\zeta(\xi)$ is periodic with period $2i$ and is free of singularities
in the strip $\vert{\rm Im\ }\xi\vert\leq1$ as can be seen from (\ref{zetaQQ}),
it is actually an entire function and thus a Laurent polynomial in the
variable $e^{\pi\xi}$. If $d<p_0$, as is the case for $\tilde{\cal A}$, then
the only such polynomial consistent with the asymptotic equation 
(\ref{asy}) is $\zeta(\xi)=c_0={\rm const.}$ Similarly for
$\tilde{\cal B}_2$ (\ref{asy}) forces $\zeta(\xi)$ to vanish.
In this latter case the $D_{p_0}$ type Y-system is further reduced
since from (\ref{zetaQQ}) we have in this case
\be
T_{p_0-1}(\xi)=0\qquad{\rm and}\qquad
Y_{p_0-2}(\xi)=0.
\end{equation}
The remaining Y-system for $\{Y_s(\xi), s=1,2,\dots,p_0-3\}$ is of type
$A_{p_0-3}$.

For later use we note that in case $\tilde{\cal A}$
\be
T_{p_0-1}(\xi)=
c_0Q(\xi+ip_0)Q(\xi-ip_0)
\label{noroot1}
\end{equation}
and this implies that $T_{p_0-1}(\xi)$ has no zeroes in the
strip $\vert{\rm Im\ }\xi\vert\leq1$. Similarly
in case $\tilde{\cal B}_2$
\be
T_{p_0-2}(\xi)=(-1)^{m+1}(-1)^{\frac{p_0}{2}} T_0(\xi+ip_0)
\label{noroot2}
\end{equation}
implies that $T_{p_0-2}(\xi)$ has no zeroes in the
strip $\vert{\rm Im\ }\xi\vert\leq1$. 

To summarize, for integer $p_0=p+1$ we found three different types of
finite Y-systems. The Y-functions can be associated with the nodes 
of the Dynkin-diagrams shown in Figures 1 and 2. 
Cases ${\cal A}$ and ${\cal B}_2$ correspond to the $D_{p+1}$-type
diagram of Figure 1$a$ and case  ${\cal B}_1$ corresponds to
the $A^s_{2p-1}$-type diagram of Figure 1$b$. The $A_{p-2}$-type
diagram of Figure 2 corresponds to the sub-case $\tilde{\cal B}_2$.
The Y-system equations are of the form 
\be
W_a(\xi+i)W_a(\xi-i)=\prod_b\left[1+W_b(\xi)\right]^{I_{ab}}.
\label{Wsystem}
\end{equation}
Here $W_a(\xi)=Y_a(\xi)$ for almost all cases with the exception
of the last two nodes of the $D_{p+1}$ diagram (Figure 1$a$). In this
case 
\be
W_{p+1}(\xi)=W_p(\xi)=K(\xi).
\end{equation}
The matrix elements $I_{ab}$ are defined as follows. $I_{ab}$ is zero
if nodes $a$ and $b$ are not connected by links and it is unity if
the nodes are connected by a single line. Finally the oriented double
line at the end of the $A^s_{2p-1}$ type diagram means
\be
I_{p-1\,\, p}=1\qquad\qquad I_{p\,\, p-1}=2.
\end{equation}

Our $D_{p+1}$ and $A_{p-2}$-type Y-system equations  
coincide with the usual
Y-system equations \cite{DTBA} associated with the Dynkin-diagrams
of these simply laced Lie algebras. Our equations corresponding to
the diagram of Figure 1$b$ correspond to those solutions of the
standard Y-system equations for the Lie algebra $A_{2p-1}$ that
are symmetric under reflection of the Dynkin diagram.

\section{The counting function}

The next step in the Bethe Ansatz solution of the model is the
definition of the \lq\lq counting" function $Z_N$ \cite{DdV0,DdV2,DdV1},
which will play an important role in our considerations.
In (\ref{exponential}) we have already defined the exponential 
counting function which
can be used to find the positions of Bethe Ansatz \lq\lq holes",
real solutions of (\ref{BAE1}), different from the Bethe roots:
\be
{\cal F}(h_\alpha)=-1\qquad\qquad \alpha=1,\dots,H.
\label{holes}
\end{equation} 

To define $Z_N$ properly, we need some more definitions. If the function
$f(z)$ is analytic and non-vanishing (ANZ) in some simply connected
domain ${\cal U}$ then we can define its \lq\lq logarithm"
${\rm log\,}f(z)$ as the primitive function of $f^\prime(z)/f(z)$,
up to a constant, which can be fixed by giving the value
of ${\rm log\,}f(z_0)$ for some $z_0\in{\cal U}$.

Using the above definition we now introduce the functions
\be
\phi_r(\xi)=-i\,{\rm log\,}\frac
{\sinh\frac{\gamma}{2}(ir-\xi)}
{\sinh\frac{\gamma}{2}(ir+\xi)}
,\qquad\qquad \phi_r(0)=0
\end{equation} 
in the strip $\vert{\rm Im}\xi\vert<r$ (for $0<r<p_0$).
For later use we note that
\be
\phi_r(\pm\infty)=\pm\pi\,\frac{p_0-r}{p_0}.
\label{infty1}
\end{equation}

The counting function $Z_N$ is now defined as\footnote{
The $\frac{\pi}{2}$ factor is present in the argument of  $Z_N$
because our variable $\xi$ differs by this factor from the usual
rapidity variable.}
\be
Z_N\left(\frac{\pi}{2}\xi\right)=\frac{N}{2}\left\{
\phi_1\left(\xi-\frac{2\Theta}{\pi}\right)
+\phi_1\left(\xi+\frac{2\Theta}{\pi}\right)\right\}-
\sum_{j=1}^m\phi_2(\xi-w_j).
\end{equation} 
The precise relation between $Z_N$ and the exponential counting
function is
\be
{\cal F}(\xi)=(-1)^\delta\,e^{iZ_N\left(\frac{\pi}{2}\xi\right)},
\end{equation} 
where
\be
\delta\equiv m+\frac{2\omega}{\pi}\qquad ({\rm mod\,} 2).
\end{equation} 
Using (\ref{infty1}) we can calculate
\be
Z_N(\pm\infty)=\pm\pi\left(m+2d-\frac{2d}{p_0}\right).
\label{infty2}
\end{equation}

In the following we will assume that the Bethe Ansatz solution
we are considering is such that the counting function is
essentially monotonically increasing. We require that
\bea
Z^\prime_N\left(\frac{\pi}{2}w_j\right)&>&0\qquad\qquad j=1,\dots,m,\\
Z^\prime_N\left(\frac{\pi}{2}h_\alpha\right)&>&0\qquad\qquad 
\alpha=1,\dots,H,\\
Z^\prime_N\left(\frac{\pi}{2}\xi\right)&>&0\qquad\qquad 
\mbox{for large $\vert\xi\vert$}
\end{eqnarray}
and call such solutions non-degenerate. We will see that all solutions
are non-degenerate for large enough $l$ but we assume that this
property of the solutions is sustained also for smaller values of $l$.

For non-degenerate solutions (\ref{infty2}) can be used to obtain
a simple formula that gives the number of holes in terms of
Bethe Ansatz data:
\bea
\omega=0:\qquad 
H&=&2d-2\left[\frac{1}{2}+\frac{d}{p_0}\right]\qquad\qquad
\delta\equiv m\ ({\rm mod\,}2),\label{count1}\\ 
\omega=\frac{\pi}{2}:\qquad 
H&=&2d-1-2\left[\frac{d}{p_0}\right]\qquad\qquad
\delta\equiv m+1\ ({\rm mod\,}2).\label{count2} 
\end{eqnarray} 
(In the above formulae the square brackets stand for integer part.)

(\ref{count1}) and (\ref{count2}) show that the $\omega=0$ sector
originally studied \cite{DdV0,DdV2} corresponds to states 
containing an even number of
holes, whereas states with odd number of holes are in the 
$\omega=\frac{\pi}{2}$ sector. The twist parameter $\omega$ has been
used in the description of restricted Sine-Gordon models \cite{DdVres} 
but here we see that it can also be used to study the odd particle sector
in the original (unrestricted) Sine-Gordon and Massive Thirring
models. The description of this important sector of the theory 
(containing the physically most interesting one-particle
states) has been based on conjectures~\cite{DdV1} but as we see here it can be
studied on an equal footing with the even sector.

\section{Lattice TBA equations}

We now translate the the Y-system functional relations (\ref{Wsystem})
into TBA integral equations \cite{JKS,Kuniba}. 
As explained in the Appendix, we need to
know the zeroes and poles of $W_a(\xi)$ in the strip
$\vert{\rm Im\,}\xi\vert<1$ together with their asymptotic behaviour.
We will call this strip the main strip.
From the definitions (\ref{Y}) and (\ref{K}) it is clear that all 
$W_a(\xi)$ are bounded at infinity and that in the main strip 
they can have zeroes only, there are no poles. It is also clear from
these definitions that the zeroes of $W_a(\xi)$ are inherited from
the neighbouring $T_s(\xi)$ functions. 

Let the set of zeroes of $T_k(\xi)$ in the main strip be 
\be
\hat r_k=\left\{\hat y_k^{(n)}\right\}_{n=1}^{\hat R_k}.
\end{equation}
The zeroes are not necessarily different from each other, for example
\be
\hat r_0=\frac{N}{2}\left\{
\frac{2\Theta}{\pi}, -\frac{2\Theta}{\pi}\right\},\qquad\qquad
\hat R_0=N.
\end{equation}

The zeroes of $T_1(\xi)$ are solutions of the equation
\be
{\cal F}(\hat y^{(n)}_1)=-1,
\label{HOLES}
\end{equation}
different from the Bethe roots $w_j$. This set includes the holes
defined in (\ref{holes}) and may contain $C$ additional complex \lq\lq
holes", complex solutions of (\ref{HOLES}) in the main strip.
Thus, $\hat R_1=H+C$.

The set of zeroes of $W_a(\xi)$ will be denoted by
\be
\hat q_a=\left\{\hat z_a^{(\alpha)}\right\}_{\alpha=1}^{\hat Q_a}
\end{equation}
and is given by
\bea
D_{p+1}:\qquad \hat q_a&=&\hat r_{a-1}\cup\hat r_{a+1}\qquad
a=1,\dots,p-1\\ 
\hat q_p&=&\hat q_{p+1}=\hat r_{p-1}\\ 
\bigskip
A^s_{2p-1}:\qquad \hat q_a&=&\hat r_{a-1}\cup\hat r_{a+1}\qquad
a=1,\dots,p-1\\ 
\hat q_p&=&\hat r_{p-1}\cup\hat r_{p+1}=2\hat r_{p-1}\\ 
\bigskip
A_{p-2}:\qquad \hat q_a&=&\hat r_{a-1}\cup\hat r_{a+1}\qquad
a=1,\dots,p-3\\ 
\hat q_{p-2}&=&\hat r_{p-3}\label{last} 
\end{eqnarray}
We note that in (\ref{last}) we have used the fact that $\hat
R_{p-1}=0$ as established after (\ref{noroot2}) and that in the
special case $\tilde{\cal A}$ $\hat R_p=0$ as follows from 
(\ref{noroot1}).

To be able to use the results of the Appendix we make the following
definitions.
\be
l_a(u)=\sum_b\,I_{ab}\ln[1+W_b(u)],
\end{equation}
\be
\beta_a(\xi)=\frac{1}{4}\,\int_{-\infty}^\infty
du\,\frac{l_a(u)}{\cosh\frac{\pi}{2}(\xi-u)},
\label{betaa}
\end{equation}
\be
\alpha_a(u)=\frac{1}{4}\,{\cal P}\,\int_{-\infty}^\infty
dv\,\frac{l_a(v)}{\sinh\frac{\pi}{2}(v-u)}.
\label{alphaa}
\end{equation}
(In the last formula ${\cal P}$ stands for principal value integration.)

Using the main result (\ref{solFR}) of the Appendix we now transform the 
Y-system equations (\ref{Wsystem}) into TBA integral equations:
\be
W_a(\xi)=\hat\sigma_a\,\prod_{\alpha=1}^{\hat Q_a}\,
\tau\left(\xi-\hat z^{(\alpha)}_a\right)\,\exp\left\{
\beta_a(\xi)\right\},
\label{latticeTBA}
\end{equation}
where $\hat\sigma_a$ is the sign of $W_a(\infty)$ and $\tau(\xi)$
is defined in (\ref{tau}).

The integral equations (\ref{latticeTBA}) are not sufficient to reconstruct
the functions $W_a(\xi)$ completely. They have to be supplemented by
the \lq\lq quantization conditions"
\be
1+W_a\left(\hat y^{(n)}_a\pm i\right)=0
\qquad n=1,\dots,\hat R_a\qquad a=1,\dots,p.
\label{latticeQC}
\end{equation}
($a=1,\dots,p-2$ for the $A_{p-2}$ system.)

It is easy to see that (\ref{latticeQC}) follows from the Y-system
equations but it is also true that (\ref{latticeTBA})
and (\ref{latticeQC}) together are completely equivalent to the Y-system
equations (\ref{Wsystem}).

\section{Continuum limit}

We are interested in the finite volume continuum limit, which is obtained by
taking $N\to\infty$ and tuning the inhomogeneity parameter
simultaneously according to (\ref{Theta}). In the Bethe Ansatz
language in this limit we have to take $m\to\infty$ while keeping
$d, H$ fixed.

We assume that the Y-system functions $W_a(\xi)$ have well-defined
continuum limits. The corresponding limit of the set of zeroes 
$\hat r_k$ will be denoted by $r_k$:
\be
\hat r_k=\left\{\hat y^{(n)}_k\right\}_{n=1}^{\hat R_k}
\quad\longrightarrow\quad r_k=\left\{y^{(n)}_k\right\}_{n=1}^{R_k},
\end{equation}
where $R_k\leq\hat R_k$, since some of the zeroes can disappear
in the process by going to infinity. 

Actually, all the zeroes of $T_0(\xi)$ go away in the limit and the 
corresponding factor
\be
C_N(\xi)=\prod_{n=1}^N\tau\left(\xi-\hat y_0^{(n)}\right) 
= (-1)^{\frac{N}{2}}\left\{
\frac{1-\frac{l}{2N}e^{\frac{\pi}{2}\xi}}
{1+\frac{l}{2N}e^{\frac{\pi}{2}\xi}}\,\,
\frac{1-\frac{l}{2N}e^{-\frac{\pi}{2}\xi}}
{1+\frac{l}{2N}e^{-\frac{\pi}{2}\xi}}\right\}^{\frac{N}{2}}
\end{equation}
becomes
\be
(-1)^{\frac{N}{2}}\,e^{-l\cosh\frac{\pi}{2}\xi}
\end{equation}
in the continuum limit.

We denote by $\sigma_a$ the sign of $W_a(\infty)$ in the continuum limit, 
which can also differ from $\hat\sigma_a$.

In the continuum limit
\be
\hat q_a\quad\longrightarrow\quad q_a=
\left\{z_a^{(\alpha)}\right\}_{\alpha=1}^{Q_a},
\end{equation}
where
\be
q_1=r_2,\qquad 
q_a=r_{a-1}\cup r_{a+1}
\qquad\qquad a=2,\dots,p-1
\end{equation}
($a=2,\dots,p-3$ for the $A_{p-2}$ case) and
\bea
q_p=q_{p+1}&=&r_{p-1}\ \ \ \ \ \ \ (D_{p+1}),\nonumber\\
q_p&=&2\cdot r_{p-1}\ \ \ \,(A_{2p-1}^s),\\
q_{p-2}&=&r_{p-3}\ \ \ \ \ \ \ (A_{p-2}\ \ p\geq5).\nonumber
\end{eqnarray}

With these definitions the TBA integral equations in the continuum
limit become
\be
W_a(\xi)=\sigma_a\,e^{-\delta_{a1}\,l\cosh\frac{\pi}{2}\xi}
\prod_{\alpha=1}^{Q_a}\,
\tau\left(\xi-z^{(\alpha)}_a\right)\,\exp\left\{
\beta_a(\xi)\right\},
\label{conTBA}
\end{equation}
which have to be supplemented by the quantization conditions
\be
1+W_a\left(y^{(n)}_a\pm i\right)=0
\qquad n=1,\dots,R_a\qquad a=1,\dots,p.
\label{conQC}
\end{equation}
($a=1,\dots,p-2$ for the $A_{p-2}$ system.)
The exponential factor $e^{-l\cosh\frac{\pi}{2}\xi}$ is present
in the TBA equation for $a=1$ only. This is indicated
in Figures 1 and 2 by colouring the corresponding nodes black.

An important special case is when all zeroes are real. In this case
the modulus of $W_a\left(y^{(n)}_a\pm i\right)$ is automatically
equal to unity and (\ref{conQC}) can be rewritten as
\be
(i)^{Q_a}\,
\exp-i\left\{\delta_{a1}l\sinh\left(\frac{\pi}{2}y^{(n)}_a\right)
-\alpha_a\left(y^{(n)}_a\right)+
\sum_{\alpha=1}^{Q_a}\gamma\left(y^{(n)}_a-z^{(\alpha)}_a\right)
\right\}=-\sigma_a,
\label{realQC}
\end{equation}
for $n=1,\dots,R_a$ and $a=1,\dots,p$ ($a=1,\dots,p-2$ for $A_{p-2}$).
In (\ref{realQC}) the notation
\be
\gamma(u)=2\arctan\big(\tau(u)\big)
\end{equation}
is used. Note that $\vert\gamma(u)\vert\leq\frac{\pi}{2}$ for real $u$.

\section{Energy and momentum}

The formulae (\ref{EP}) determinig the energy and momentum of the Bethe
Ansatz state can be rewritten in terms of $T_0(\xi)$ and $T_1(\xi)$ as
\be
e^{-2iaE}=\frac
{T_1\left(\frac{2\Theta}{\pi}-i\right)
T_1\left(i-\frac{2\Theta}{\pi}\right)}
{T_0^2\left(\frac{2\Theta}{\pi}-2i\right)},\qquad\qquad
e^{-2iaP}=e^{-4i\omega}\,\,\frac
{T_1\left(\frac{2\Theta}{\pi}-i\right)}
{T_1\left(i-\frac{2\Theta}{\pi}\right)}.
\label{EP2}
\end{equation}

From (\ref{1Y}) we see that $T_1(\xi)$ satisfies
\be
T_1(\xi+i)T_1(\xi-i)=\left[1+Y_1(\xi)\right]
T_0(\xi+2i)T_0(\xi-2i),
\label{T1Y1}
\end{equation}
which is of the form (\ref{fr}) and can be solved by the techniques
explained in the Appendix. For this purpose we first define
\be
T_1(\xi)=\hat\sigma\,\tilde T_1(\xi)\,\Phi^{N/2}(\xi),
\label{tildeT1}
\end{equation}
where $\hat\sigma$ is a sign, $\tilde T_1(\xi)$ has the same zeroes as
$T_1(\xi)$, has no poles, it is positive for $\xi\to\infty$, and satisfies
\be
\tilde T_1(\xi+i)\tilde T_1(\xi-i)=1+Y_1(\xi).
\label{tildeT1Y1}
\end{equation}
In (\ref{tildeT1}) the factor $\Phi(\xi)^{N/2}$ solves the
$T_0$-dependent part of (\ref{T1Y1}):
\be
\Phi(\xi)=\exp\left\{\Delta(\xi)\right\},\qquad\qquad
\Delta(\xi)=\frac{1}{4}\,\int_{-\infty}^\infty
du\,\frac{\ln B\left(u,\frac{2\Theta}{\pi}\right)}
{\cosh\frac{\pi}{2}(\xi-u)},
\end{equation}
where
\be
B(u,\lambda)=k(u+\lambda+2i)\,k(u+\lambda-2i)\,
k(u-\lambda+2i)\,k(u-\lambda-2i).
\end{equation}

Now we take the logarithm of (\ref{EP2}) and get
\bea
E&=&E_0+\bar E\qquad\qquad
\bar E=S_++S_-,\\
P&=&P_0+\bar P\qquad\qquad
\bar P=S_+-S_-,
\end{eqnarray}
where
\be
S_+=\frac{i}{2a}\ln\tilde T_1\left(\frac{2\Theta}{\pi}-i\right)
\qquad\qquad
S_-=\frac{i}{2a}\ln\left[(-1)^H
\tilde T_1\left(i-\frac{2\Theta}{\pi}\right)\right]
\label{Spm}
\end{equation}
and the energy and momentum constants are
\bea
E_0&=&\frac{\pi}{a}N_0-\frac{\omega}{a}+\frac{iN}{2a}
\left\{\Delta\left(\frac{2\Theta}{\pi}-i\right)-\frac{i\pi}{2}+
\ln\sin\gamma-\ln\sinh\gamma\left(\frac{2\theta}{\pi}-i\right)
\hspace{-1.5mm}\right\}
\hspace{-1.5mm},\\
P_0&=&\frac{\omega}{a}+\frac{\pi}{a}N_1.
\end{eqnarray}
Here the choice of the integers $N_0$ and $N_1$ (which are in principle 
arbitrary) is part of the regularization scheme. If we choose $N_0=N_1=0$
we get
\be
P_0=\frac{\omega}{a}\qquad\qquad{\rm and}\qquad\qquad
E_0=-\frac{\omega}{a}+\frac{N}{2a}\chi_0\left(\frac{4\Theta}{\pi}\right).
\label{E0P0}
\end{equation}
The precise definition of the function $\chi_0(\xi)$ will be given
later. What is important here is that $E_0$ and $P_0$ are universal
in the sense that they are the same for all states in a given sector
characterized by $\omega=0$ or $\omega=\frac{\pi}{2}$. Omitting these
unphysical constants leaves us with the physical energy and momentum
eigenvalues $\bar E$ and $\bar P$.

Using the method outlined in the Appendix we can solve (\ref{tildeT1Y1}):
\be
\tilde T_1(\xi)=
\prod_{\alpha=1}^H\tau(\xi-h_\alpha)\,
\prod_{\beta=1}^C\tau(\xi-\Omega_\beta)\,
\exp\left\{\frac{1}{4}\,\int_{-\infty}^\infty
du\,\frac{\ln[1+Y_1(u)]}
{\cosh\frac{\pi}{2}(\xi-u)}\right\}.
\end{equation}
Here $\Omega_\beta$ are complex solutions of (\ref{HOLES}).
Using the definitions (\ref{Spm}) finally we get in the continuum limit
\bea
\bar E&=&
{\cal M}\left[\sum_{\alpha=1}^H\cosh\frac{\pi h_\alpha}{2}+
\sum_{\beta=1}^C\cosh\frac{\pi\Omega_\beta}{2}
-\frac{1}{4}\int_{-\infty}^\infty\,du\cosh\frac{\pi u}{2}
\ln[1+Y_1(u)]\right]\hspace{-1.5mm},\label{ener}\\
\bar P&=&
{\cal M}\left[\sum_{\alpha=1}^H\sinh\frac{\pi h_\alpha}{2}+
\sum_{\beta=1}^C\sinh\frac{\pi\Omega_\beta}{2}
-\frac{1}{4}\int_{-\infty}^\infty\,du\sinh\frac{\pi u}{2}
\ln[1+Y_1(u)]\right]\hspace{-1mm}.\label{mom}
\end{eqnarray}

\section{The $A_1$ case}

One of the simplest cases is $p=3$, $d=H=1$. In this case we can
determine the energy and momentum eigenvalues exactly.

In this case $\delta\equiv\frac{N}{2}$ and from (\ref{Y}) it follows
that $Y_1(\infty)>0$. Further $\beta_1(\xi)=0$ and $\hat R_2=0$ here
and (\ref{latticeTBA}) reduces to
\be
Y_1(\xi)=W_1(\xi)=C_N(\xi).
\end{equation}

It is easy to see that there cannot be any complex holes in this case.
Thus $\hat R_1=1$ and we introduce $\hat y_1^{(1)}=\hat h$ for
the position of the single (real) hole. The solution of (\ref{latticeQC})
gives
\be
\hat h=h+\frac{h_2}{N^2}+{\cal O}\left(N^{-3}\right),
\end{equation} 
where $h$ is the solution of
\be
l\sinh\frac{\pi}{2}h=2\pi M_1,\qquad\qquad M_1\equiv\frac{1+\delta}{2}
\ ({\rm mod}\,1)
\end{equation} 
and
\be
h_2=\frac{l^2}{6\pi}\,\,\frac{\sinh\left(\frac{3}{2}\pi h\right)}
{\cosh\left(\frac{1}{2}\pi h\right)}.
\end{equation} 

In the continuum limit we get
\be
Y_1(\xi)=(-1)^\delta\, e^{-l\cosh\frac{\pi}{2}\xi}
\end{equation} 
and the energy and momentum eigenvalues are
\bea
\bar E&=&
{\cal M}\cosh\frac{\pi h}{2}
-\frac{{\cal M}}{4}\int_{-\infty}^\infty\,du\cosh\frac{\pi u}{2}
\ln[1+(-1)^\delta\, e^{-l\cosh\frac{\pi}{2}u}],\label{Ea1}\\
\bar P&=&
{\cal M}\sinh\frac{\pi h}{2}=\frac{2\pi M_1}{L}.\label{Pa1}
\end{eqnarray}

It is interesting to note that the momentum eigenvalues (\ref{Pa1})
are exactly the same as they would be in a free theory. The $p=3$
model is not free however since the SG S-matrix is non-trivial at
$\beta^2=6\pi$. (The free fermion point is $\beta^2=4\pi$ or $p=1$.)  
The energy eigenvalues differ from those
of the free theory by the second term in (\ref{Ea1}). On the other
hand, this shift is independent of the momentum quantum number $M_1$.
We note that it is only the one-particle states that can be
exactly solved for $p=3$.
We also note that it is not obvious in the DdV language that $p=3, H=1$
is a special case and we were able to solve the corresponding DdV
equations only numerically.

\section{Destri-deVega equation}

In the Sine-Gordon (MT) model a non-linear integral equation
(Destri-deVega equation) is used to describe the Bethe Ansatz states
and calculate their energy and momentum \cite{DdV0,DdV2,DdV1}. 
This is formulated
in terms of the counting function $Z_N$ and its exponential, ${\cal F}(\xi)$.

Let us chose the parameter $B$ so that there are no complex holes
in the strip $0<{\rm Im}\,\xi<B$. Then we can define
\be
{\cal L}_+(\xi)={\rm log}\,\left[1+{\cal F}(\xi)\right]\qquad\qquad
0<{\rm Im}\,\xi<B.
\end{equation}
We define further the function $\chi(\xi)$ by
\be
\chi^\prime(\xi)=G(\xi),\qquad\qquad \chi(0)=0,
\end{equation}
where
\be
G(\xi)=\int_{-\infty}^\infty dk\,e^{-ik\xi}\,
\frac{\sinh(p-1)k}{2\cosh k\,\sinh pk}.
\end{equation}
The function $\chi_0(\xi)$ occurring in (\ref{E0P0}) is defined
analogously with $p$ replaced by\break $p_0=p+1$.

For non-degenerate states
\be
{\cal L}_+(\xi)=\hat L_+(\xi)=\ln\left[1+(-1)^\delta
e^{iZ_N\left(\frac{\pi}{2}\xi\right)}\right]
\end{equation}
and the DdV equation can be written as
\begin{gather}
Z_N\left(\frac{\pi}{2}\xi\right)=\frac{N}{2}\Big\{
\arctan\left[\sinh\left(\frac{\pi}{2}\xi-\Theta\right)\right]
+\arctan\left[\sinh\left(\frac{\pi}{2}\xi+\Theta\right)\right]\Big\}
+\sum_{\alpha=1}^H\chi(\xi-h_\alpha)\notag\\
+\frac{1}{2\pi i}\int_{-\infty}^\infty\,du\big[
G(\xi-u-i\eta)\hat L_+(u+i\eta)-G(\xi-u+i\eta)\hat L_+^*(u+i\eta)
\big],
\end{gather}
where $0<\eta<B$.
Analogously to the quantization conditions (\ref{latticeQC}) supplementing
the TBA integral equations the DdV equations are supplemented
by the quantization conditions
\be
Z_N\left(\frac{\pi}{2}h_\alpha\right)=2\pi M_\alpha\qquad\qquad
M_\alpha\equiv\frac{1+\delta}{2}\ ({\rm mod}\,1)\qquad\qquad
\alpha=1,\dots,H.
\label{DdVQC}
\end{equation}

The counting function $Z_N$ has continuum limit $Z$ in terms of which
the continuum DdV equation reads
\begin{gather}
Z\left(\frac{\pi}{2}\xi\right)=l\sinh\frac{\pi}{2}\xi
+\sum_{\alpha=1}^H\chi(\xi-h_\alpha)\qquad\qquad\notag\\
+\frac{1}{2\pi i}\int_{-\infty}^\infty\,du\big[
G(\xi-u-i\eta)L_+(u+i\eta)-G(\xi-u+i\eta)L_+^*(u+i\eta)
\big],\label{DdV}
\end{gather}
where
\be
L_+(\xi)=\ln\left[1+(-1)^\delta
e^{iZ\left(\frac{\pi}{2}\xi\right)}\right].
\label{Lplus}
\end{equation}

The energy and momentum in the continuum can be expressed in terms
of the positions of the holes $h_\alpha$ and $L_+$:
\bea
\bar E&=&
{\cal M}\sum_{\alpha=1}^H\cosh\frac{\pi h_\alpha}{2}
-\frac{{\cal M}}{2}\,{\rm Im}\,
\int_{-\infty}^\infty\,du\sinh\frac{\pi}{2}(u+i\eta)\,
L_+(u+i\eta),\\
\bar P&=&
{\cal M}\sum_{\alpha=1}^H\sinh\frac{\pi h_\alpha}{2}
-\frac{{\cal M}}{2}\,{\rm Im}\,
\int_{-\infty}^\infty\,du\cosh\frac{\pi}{2}(u+i\eta)\,
L_+(u+i\eta).
\end{eqnarray}

For large physical size $l$ the function $L_+$ is exponentially small
and the integrals containing it in the DdV equation (\ref{DdV}) and
also in the energy and momentum expressions can be neglected. In this 
approximation the quantization conditions (\ref{DdVQC}) are nothing
but the (dressed) Bethe Ansatz equations for solitons/fermions
in the SG/MT model. This leads to the identification
\bea
{\rm SG:}\ \ ({\rm soliton-soliton})\ \ &&S(\theta)=-\exp\left\{
i\chi\left(\frac{2\theta}{\pi}\right)\right\}\label{SG}\\
&&\delta\equiv H\quad({\rm mod}\,2)\notag
\end{eqnarray}
and
\bea
{\rm MT:}\ \ ({\rm fermion-fermion})\ \ &&S(\theta)=\exp\left\{
i\chi\left(\frac{2\theta}{\pi}\right)\right\}\label{MT}\\
&&\delta=0,\notag
\end{eqnarray}
where $S(\theta)$ is the scattering phase for soliton/fermion scattering.

The parameter $p$ is related to the usual SG coupling $\beta$ by
\be
\beta^2=\frac{8\pi p}{p+1}.
\end{equation}
$p=1$ corresponds to the free fermion point and $p=\infty$ to the
XY model (asymptotically free point).

For large $l$ the first term on the right hand side of (\ref{DdV})
dominates and this shows that $Z\left(\frac{\pi}{2}\xi\right)$ 
is monotonically increasing for real $\xi$. We already used the
assumption that this property remains valid also for smaller $l$.

The DdV equation (\ref{DdV}) is suitable for numerical calculations.
To study the analytic properties of the solution it is useful
to rewrite it in the form
\be
Z\left(\frac{\pi}{2}\xi\right)=l\sinh\frac{\pi}{2}\xi
+\sum_{\alpha=1}^H\chi(\xi-h_\alpha)
+\int_{-\infty}^\infty\,du\, G(\xi-u)Q(u),
\label{DdVQ}
\end{equation}
where
\be
Q(u)=\frac{1}{\pi}\lim_{\eta\to0}\,{\rm Im}\,L_+(u+i\eta).
\end{equation}
The representation (\ref{DdVQ}) is valid in the strip 
$\vert {\rm Im}\,\xi\vert<2$. Note that we are considering only
the repulsive case $p>1$ in this paper. In the attractive regime
$p<1$ some of our expressions have to be modified.

We now analytically continue the counting function to the whole
complex $\xi$ plane. For this purpose we define
\bea
\gamma_1(\xi)=-i\,{\rm log}\,w(\xi)\ \ \ \qquad\gamma_1(ip)&=&\pi\qquad
\ \ \ \ \ 1<{\rm Im}\,\xi<2p-1,\\
\gamma_2(\xi)=-i\,{\rm log}\,w(\xi)\,\qquad\gamma_2(-ip)&=&\pi\qquad
-2p+1<{\rm Im}\,\xi<-1,
\end{eqnarray}
where
\be
w(\xi)=\frac
{\sinh\frac{\pi}{2p}(i-\xi)}
{\sinh\frac{\pi}{2p}(i+\xi)}.
\end{equation}

The analytical continuation of the counting function in the strip
$2<{\rm Im}\,\xi<2p$ is
\be
Z_1\left(\frac{\pi}{2}\xi\right)=\delta\pi
+\sum_{\alpha=1}^H\gamma_1(\xi-i-h_\alpha)
+\int_{-\infty}^\infty\,du\, K_1(\xi-u)Q(u),
\label{DdVQ1}
\end{equation}
where
\be
K_1(\xi)=G(\xi)+G(\xi-2i),
\end{equation}
which is analytic in this strip.

Similarly in the strip $-2p<{\rm Im}\,\xi<-2$ the analytical
continuation is
\be
Z_2\left(\frac{\pi}{2}\xi\right)=\delta\pi
+\sum_{\alpha=1}^H\gamma_2(\xi+i-h_\alpha)
+\int_{-\infty}^\infty\,du\, K_2(\xi-u)Q(u),
\label{DdVQ2}
\end{equation}
where
\be
K_2(\xi)=G(\xi)+G(\xi+2i),
\end{equation}
which is analytic in this strip.

The exponential counting function is thus defined as
\be
{\cal F}(\xi)=\left\{
\begin{array}{lc}
(-1)^\delta\,e^{iZ_1\left(\frac{\pi}{2}\xi\right)}&
\qquad\qquad 2<{\rm Im}\,\xi<2p\\
(-1)^\delta\,e^{iZ\left(\frac{\pi}{2}\xi\right)}&
\qquad\qquad \vert{\rm Im}\,\xi\vert<2\\
(-1)^\delta\,e^{iZ_2\left(\frac{\pi}{2}\xi\right)}&
\qquad\qquad -2p<{\rm Im}\,\xi<-2
\end{array}
\right.
\end{equation}
and it is easy to see that this defines a meromorphic function 
with properties
\be
{\cal F}(\xi+2ip_0)={\cal F}(\xi),\qquad\qquad
{\cal F}^*(\xi)=\frac{1}{{\cal F}(\xi^*)}.
\end{equation}
It has poles ($\infty$ many) at $\xi=w_j+2i$ and zeroes ($\infty$
many) at $\xi=w_j-2i$.

\section{Infinite volume solution}

We have seen that for large $l$ the integrals give exponentially
small contribution to the counting function $Z$. This is also true for 
$Z_1$ and $Z_2$ although this cannot be seen immediately from (\ref{DdVQ1})
and (\ref{DdVQ2}). However, it is possible to write $Z_1$ and $Z_2$
in a form analogous to (\ref{DdV}) and show that the corresponding
integrals are indeed exponentially small. Neglecting the integrals we have
 \be
{\cal F}(\xi)=\left[1+{\cal O}\left(e^{-{\cal K}l}\right)\right]\cdot\left\{
\begin{array}{lc}
u(\xi)u(\xi-2i)&
\qquad\qquad 2<{\rm Im}\,\xi<2p\\
e^{il\sinh\frac{\pi}{2}\xi}\,u(\xi)&
\qquad\qquad \vert{\rm Im}\,\xi\vert<2\\
u(\xi)u(\xi+2i)&
\qquad\qquad -2p<{\rm Im}\,\xi<-2
\end{array}
\right.
\end{equation}
where ${\cal K}$ is ${\rm Im}\,\xi$-dependent and positive.
The function $u(\xi)$ is defined by
\be
u(\xi)=(-1)^\delta\,\prod_{\alpha=1}^H\sigma(\xi-h_\alpha),
\label{u}
\end{equation}
where
\be
\sigma(\xi)=e^{i\chi(\xi)}
\end{equation}
is meromorphic with poles and zeroes on the imaginary axis and satisfies
\be
\sigma(\xi+i)\sigma(\xi-i)=w(\xi).
\end{equation}

Since we can express the Y-system functions in terms of the counting
function starting with (\ref{FY}) we are able to calculate also the Y-functions
with exponential precision for large $l$. For $Y_1(\xi)$ we find
 \be
Y_1(\xi)=\left[1+{\cal O}\left(e^{-{\cal K}l}\right)\right]\cdot\left\{
\begin{array}{lc}
\eta_2(\xi-i)&
\qquad\qquad 3<{\rm Im}\,\xi<2p-1\\
\lambda(\xi)e^{-l\cosh\frac{\pi}{2}\xi}&
\qquad\qquad \vert{\rm Im}\,\xi\vert<3\\
\eta_2(\xi+i)&
\qquad\qquad -2p+1<{\rm Im}\,\xi<-3
\end{array}
\right.
\label{Y1asy}
\end{equation}
and using the Y-system equations (\ref{Ysystem}) recursively we get
for $k=2,\dots,p$ 
\be
Y_k(\xi)=\left[1+{\cal O}\left(e^{-{\cal K}l}\right)\right]
\eta_k(\xi)
\qquad\qquad \vert{\rm Im}\,\xi\vert<k.
\label{Ykasy}
\end{equation}
In (\ref{Y1asy})
\be
\lambda(\xi)=u(\xi+i)+\frac{1}{u(\xi-i)}
\label{lam}
\end{equation}
and
\be
\eta_2(\xi)=\lambda(\xi+i)\lambda(\xi-i)-1.
\label{eta2}
\end{equation}
The functions $\eta_k(\xi)$ satisfy the Y-system equations 
\be
\eta_k(\xi+i)\eta_k(\xi-i)=\left[1+\eta_{k+1}(\xi)\right]
\left[1+\eta_{k-1}(\xi)\right]
\label{etasystem}
\end{equation}
for $k=2,3,\dots$ and this determines\footnote{$\eta_1(\xi)=0$ by definition.} 
$\eta_k(\xi)$ for $k>2$.

It is possible to find the solution of (\ref{etasystem}) explicitly.
We note first that there is a class of solutions depending on
a parameter $q$ and a function $B(\xi)$. Using these input data
we first define 
\be
t_0(\xi)=0,\qquad
t_k(\xi)=\sum_{j=0}^{k-1}q^j\,B\left[\xi+i(k-1-2j)\right]\qquad
k=1,2\dots
\end{equation}
and then it is easy to show that
\be
\eta_k(\xi)=q^{1-k}\,\frac{t_{k+1}(\xi)t_{k-1}(\xi)}
{B(\xi+ik)B(\xi-ik)}\qquad\qquad k=1,2,\dots
\label{eta}
\end{equation}
solve (\ref{etasystem}). (\ref{eta}) is quite analogous to 
(\ref{Y}) and it is also true that
\be
1+\eta_k(\xi)=q^{1-k}\,\frac{t_k(\xi+i)t_k(\xi-i)}
{B(\xi+ik)B(\xi-ik)}\qquad\qquad k=1,2,\dots
\label{1eta}
\end{equation}

The actual solution entering (\ref{Y1asy}) and (\ref{Ykasy}) corresponds
to the choice
\be
q=(-1)^H\qquad\qquad B(\xi)=\prod_{\alpha=1}^H\,
\sinh\frac{\pi}{2p}(\xi-h_\alpha).
\end{equation}
This can be verified by computing $\eta_2(\xi)$ from 
(\ref{eta2}) and from (\ref{eta}) with the above choice and showing
that they coincide.

Similarly to the general Y-system equations, the set of
$\eta$-functions can also be truncated. In cases ${\cal A}$ and
${\cal B}_2$ we can define
\be
\kappa(\xi)=\frac{t_{p-1}(\xi)}{B(\xi-ip)}
\end{equation}
and then the relations
\be
\kappa(\xi+i)\kappa(\xi-i)=1+\eta_{p-1}(\xi)\qquad\qquad
1+\eta_p(\xi)=\left[1+\kappa(\xi)\right]^2
\end{equation}
can be used to show that the truncated system is of type $D_{p+1}$.
If $H<p$ is also satisfied ($\tilde{\cal B}_2$ case) then
$t_p(\xi)=0$ implies $\eta_{p-1}(\xi)=0$ and this corresponds to
the further truncation to $A_{p-2}$. Finally in the ${\cal B}_1$ case
reflection relations analogous to (\ref{refl1},\ref{refl2})
are satisfied by $t_k(\xi)$ and $\eta_k(\xi)$ and the truncation is
of type $A_{2p-1}^s$ accordingly.

Similarly to the full Y-system where the zeroes are determined by 
those of the T-system, the zeroes of $t_k(\xi)$ determine the
zeroes of $\eta_k(\xi)$. More concretely,
\bea
\lambda(\xi): \hspace{-3mm} &&{\rm ANZ}\qquad\vert{\rm Im}\,\xi\vert<3
\qquad\mbox{except for zeroes of \ \ $t_2(\xi)$},\notag\\
\eta_k(\xi): \hspace{-3mm} &&{\rm ANZ}\qquad\vert{\rm Im}\,\xi\vert<k
\qquad\mbox{except for zeroes of \ \ $t_{k+1}(\xi)t_{k-1}(\xi)$},\\
\kappa(\xi): \hspace{-3mm} &&{\rm ANZ}\qquad\vert{\rm Im}\,\xi\vert<p
\qquad\mbox{except for zeroes of \ \ $t_{p-1}(\xi)$}.\notag
\end{eqnarray}
For later use we note that
\be
\lambda(\pm\infty)=2(-1)^\delta\cos\left[
\frac{\pi H}{2}\left(1-\frac{1}{p}\right)\right].
\label{lambdainfty}
\end{equation}

\noindent We now consider some simple examples.
\medskip

\noindent\fbox{$H=0$} \break
In this case we have a $D_{p+1}$ system for $p\geq2$.
From (\ref{u}) and (\ref{lam}) we get
\be
\lambda(\xi)=2(-1)^\delta,
\label{H0lam}
\end{equation}
where, according to (\ref{SG}) and (\ref{MT}) only the choice
$\delta=0$ is physical. Further
\be
B(\xi)=1,\qquad t_k(\xi)=k,\qquad\eta_k(\xi)=k^2-1,\qquad\kappa(\xi)=p-1.
\label{H0}
\end{equation}

\medskip

\noindent\fbox{$H=1$} \break
Here, according to (\ref{SG},\ref{MT}) we have two choices:
\be
\delta=1\qquad\mbox{for SG},\qquad\qquad\qquad
\delta=0\qquad\mbox{for MT}
\end{equation}
and we have an $A^s_{2p-1}$ system for $p\geq2$ even and an $A_{p-2}$
system for $p\geq3$ odd. The position of the hole is denoted by $h_1=h$.
From (\ref{u}) and (\ref{lam}) we find in this case
\be
\lambda(\xi)=(-1)^\delta\left\{e^{i\chi(\xi-h+i)}+e^{-i\chi(\xi-h-i)}\right\}
\label{H1lam}
\end{equation}
and from
\be
B(\xi)=\sinh\frac{\pi}{2p}(\xi-h)
\end{equation}
we have
\be
t_k(\xi)=\left\{
\begin{array}{ll}
\phantom{i}\,\frac
{\cos\left(\frac{k\pi}{2p}\right)}
{\cos\left(\frac{\pi}{2p}\right)}
\sinh\frac{\pi(\xi-h)}{2p} & \qquad\mbox{$k$ odd}\\
i\,\frac
{\sin\left(\frac{k\pi}{2p}\right)}
{\cos\left(\frac{\pi}{2p}\right)}
\cosh\frac{\pi(\xi-h)}{2p} & \qquad\mbox{$k$ even}
\end{array}
\right.
\label{H1}
\end{equation}

\medskip

\noindent\fbox{$H=2$} \break
Here we have a $D_{p+1}$ system for $p\geq2$ again and only 
the choice $\delta=0$ is physical. For simplicity we restrict our
attention to the symmetric case
\be
h_1=h\qquad\qquad h_2=-h,
\end{equation}
which corresponds to
\be
B(\xi)=\frac{1}{2}\cosh\frac{\pi\xi}{p}-\frac{1}{2}\cosh\frac{\pi h}{p}.
\end{equation}
Now we find
\be
\lambda(\xi)=\left\{e^{i\chi(\xi-h+i)+i\chi(\xi+h+i)}
+e^{-i\chi(\xi-h-i)-i\chi(\xi+h-i)}\right\}
\label{H2lam}
\end{equation}
and 
\be
t_k(\xi)=\frac{1}{2}\,\frac{\sin\left(\frac{k\pi}{p}\right)}
{\sin\left(\frac{\pi}{p}\right)}\,
\cosh\frac{\pi\xi}{p}
-\frac{k}{2}\,\cosh\frac{\pi h}{p}. 
\label{H2}
\end{equation}

\section{Ground state}
The ground state of the system ($H=0$) belongs to the $\tilde{\cal A}$
case and corresponds to a $D_{p+1}$ type Y-system for $p\geq2$.
The only physical choice is $\delta=0$. From (\ref{H0}) we see that
for $l\to\infty$ there are no zeroes and all $W_a(\xi)$ functions
are positive. Our assumption is that these qualitative properties of
the solution remain valid also for finite $l$ values. 
If this is true then (\ref{conTBA}) becomes
\be
W_a(\xi)=e^{-\delta_{a1}l\cosh\frac{\pi}{2}\xi}\,
\exp\left\{\beta_a(\xi)\right\}\qquad\qquad a=1,\dots,p+1
\label{H0TBA}
\end{equation}
and there are no quantization conditions in this case.
By taking the logarithm on both sides of (\ref{H0TBA}) we reproduce
the customary TBA system describing the ground state of the SG/MT
model. Noting that both $W_a(\xi)$ and $\beta_a(\xi)$ are now
even functions we find from (\ref{ener}) and (\ref{mom})
\be
P^{(0)}=0,\qquad\qquad
E^{(0)}=-\frac{{\cal M}}{4}\,\int_{-\infty}^\infty\,du\,
\cosh\frac{\pi u}{2}\,\ln\left[1+W_1(u)\right].
\label{H0enermom}
\end{equation}

\section{One-particle states}

For one-particle states both $\delta=1$ (SG model) and $\delta=0$
(MT model) are possible. Moreover, depending on the parity of $p$,
we have an $A_{2p-1}^s$ Y-system ($p\geq2$ even) or $A_{p-2}$
Y-system ($p\geq5$ odd\footnote{We have already discussed the $A_1$ case.}).
Here we only consider the SG model with $h_1=0$ in detail, because in
this case the quantization conditions are satisfied automatically, 
which simplifies the problem enormously.

Our main assumption will be again that the qualitative features
of the $l\to\infty$ solution are also valid for finite $l$.
From (\ref{H1lam}) and (\ref{H1}) we see that
\begin{itemize}
\item $W_a(\xi):$ \ \  no zeroes for $a$ odd
\item[{}] $\phantom{W_a(\xi):}$ \ \ double zero at $\xi=0$ for $a$ even
\item all $W_a(\xi)$ are negative, except $W_p(\xi)$ in the
$A_{2p-1}^s$ case
\end{itemize}
and this leads to the TBA equations
\be
W_a(\xi)=(-1)^{1+\delta_{pa}}\,e^{-\delta_{a1}l\cosh\frac{\pi}{2}\xi}\,
\left[\tau(\xi)\right]^{1+(-1)^a}\,
\exp\left\{\beta_a(\xi)\right\}\qquad a=1,\dots,p.
\label{H1TBA}
\end{equation}
($a=1,\dots,p-2$ for $A_{p-2}$.)
$W_a(\xi)$ and $\beta_a(\xi)$ are even functions and $\alpha_a(\xi)$
are odd. It follows that the quantization conditions (\ref{realQC})
are automatically satisfied.

Finally from (\ref{ener}) and (\ref{mom}) we get
\be
P^{(1)}=0,\qquad\qquad
E^{(1)}={\cal M}-\frac{{\cal M}}{4}\,\int_{-\infty}^\infty\,du\,
\cosh\frac{\pi u}{2}\,\ln\left[1+W_1(u)\right].
\label{H1enermom}
\end{equation}

\section{Two-particle states}

The two-particle states are the only ones among our examples where
the quantization conditions (\ref{realQC}) give nontrivial constraints.
The $H=2$ states belong to the $\tilde{\cal A}$ case and correspond to
a $D_{p+1}$ type Y-system ($p\geq2$) and $\delta=0$.

Our conjecture is based again on the large $l$ solution.
For simplicity we consider only the following symmetric distribution
of zeroes:
\be
r_k=\left\{H_k,-H_k\right\}\qquad\qquad k=1,\dots,p-1,
\label{H2zeroes}
\end{equation}
where by convention $H_k>0$, the positions of the holes are $h_1=H_1$
and $h_2=-H_1$ and the signs at infinity are
\be
\sigma_1=-1,\qquad\quad \sigma_a=1\quad a=2,\dots,p-2,\qquad\quad
\sigma_{p-1}=\sigma_p=\sigma_{p+1}=-1.
\label{H2signs}
\end{equation}
It is important to note that (\ref{H2signs}) and all subsequent
formulae are valid for $p\geq3$ only. The important special case
$p=2$ will be discussed at the end of this section separately.

The TBA integral equations are
\bea
W_1(\xi)&=&-e^{-l\cosh\frac{\pi}{2}\xi}\,\tau(\xi-H_2)\tau(\xi+H_2)
\exp\left\{\beta_1(\xi)\right\},\\
W_a(\xi)&=&
\tau(\xi-H_{a-1})\tau(\xi+H_{a-1})
\tau(\xi-H_{a+1})\tau(\xi+H_{a+1})
\exp\left\{\beta_a(\xi)\right\}\hspace{-1.5mm},\label{H2Wa}\\
W_{p-1}(\xi)&=&-
\tau(\xi-H_{p-2})\tau(\xi+H_{p-2})
\exp\left\{\beta_{p-1}(\xi)\right\},\\
W_p(\xi)&=& W_{p+1}(\xi)=-
\tau(\xi-H_{p-1})\tau(\xi+H_{p-1})
\exp\left\{\beta_p(\xi)\right\}.
\end{eqnarray}
In (\ref{H2Wa}) $a=2,\dots,p-2$ and because of the symmetric
distribution of zeroes the functions $W_a(\xi)$ and $\beta_a(\xi)$
are even and $\alpha_a(\xi)$ are odd. For the same reason it is
sufficient to require that (\ref{realQC}) are satisfied for
the positive zeroes:
\bea
l\sinh\frac{\pi}{2}H_1+\gamma(H_1-H_2)+\gamma(H_1+H_2)-\alpha_1(H_1)&=&
2\pi M_1,\\
\gamma(H_a-H_{a-1})+\gamma(H_a+H_{a-1})&\phantom{=}&\nonumber\\
+\,\gamma(H_a-H_{a+1})+\gamma(H_a+H_{a+1})-\alpha_a(H_a)&=&
2\pi M_a,\label{H2Ma}\\
\gamma(H_{p-1}-H_{p-2})+\gamma(H_{p-1}+H_{p-2})-\alpha_{p-1}(H_{p-1})&=&
2\pi M_{p-1}.
\end{eqnarray}
In (\ref{H2Ma}) $a=2,\dots,p-2$ and all the quantum numbers $M_a$ are
half-integers. The value of $M_1$ is part of the input data, whereas
all others ($M_a$ for $a=2,\dots,p-1$) are determined from
consistency. Numerically we found in all cases we considered
that $M_a=1/2$ for $a=2,\dots,p-1$. Finally the two-particle
momentum and energy are
\be
P^{(2)}=0,\qquad
E^{(2)}=2{\cal M}\cosh\frac{\pi}{2}H_1-
\frac{{\cal M}}{4}\,\int_{-\infty}^\infty\,du\,
\cosh\frac{\pi u}{2}\,\ln\left[1+W_1(u)\right].
\label{H2enermom}
\end{equation}

We end this section by the discussion of the special case $p=2$ 
corresponding to the supersymmetric point of the SG model. Here the 
TBA equations are
\bea
W_1(\xi)&=&e^{-l\cosh\frac{\pi}{2}\xi}\exp\left\{\beta_1(\xi)\right\},\\
W_2(\xi)&=&W_3(\xi)=
-\tau(\xi-H_1)\tau(\xi+H_1)
\exp\left\{\beta_2(\xi)\right\}
\end{eqnarray}
and there is just one  quantization condition:
\be
l\sinh\frac{\pi}{2}H_1-\alpha_1(H_1)=2\pi M_1.
\label{}
\end{equation}

\section{Numerical iteration of the TBA equations}

We have calculated the energy and momentum eigenvalues of the ground
state and also of one- and two-particle states from the TBA equations
numerically and compared the results with those obtained using the
DdV equations. We have found agreement to many digits in all cases we 
considered and concluded that our TBA equations are completely
equivalent to the DdV equations.  

Our numerical checks covered a rather wide range of the physical
size $l$ (from ${\cal O}(1)$ down to $l\sim10^{-2}$) and we considered
a number of different choices for the parameter $p$, including
the limit $p=\infty$. The agreement between TBA and DdV is especially
spectacular for the $A_1$ case where we have exact results on the
TBA side.

The infinite volume solution we found in Section~10, which 
we have used to help postulating qualitative properties of the
Y-system elements, serve also as starting point of the numerical 
iteration procedure. We found that the convergence of the numerical 
iteration is rather fast and already the starting point is a
surprisingly good approximation not only for large volumes but also
for smaller $l$ values. In the $H=2$ case this is true in particular 
for the positions of zeroes, which change very little during the 
iteration.

\section{Summary}

In this paper we proposed TBA systems describing multi-soliton states
of the Sine-Gordon (massive Thirring) model. We derived the TBA equations
starting from the Bethe Ansatz solution of the light-cone lattice
formulation of the model, which allows a uniform treatment of states with 
even an odd soliton number. The derivation also provides a simple
link between the counting function of the Destri-deVega equation
and the Y-system component functions of the TBA approach.

After considering the continuum limit of the lattice model, the only
input we need is the distribution of zeroes of the TBA Y-functions
(and their sign at infinity). This information can be read off from
the infinite volume solution of the problem, which we found
explicitly. Our main assumption in this paper is that what we
find for the infinite volume solution remains qualitatively valid
for all physical volumes. We verified this assumption numerically
by comparing the TBA results with those obtained by the alternative  
Destri-deVega nonlinear integral equation.

The simple pattern of the excited state TBA systems found here gives us 
some hope that it will be possible to find TBA equations for excited 
states also in other models, even if no DdV type alternative is available
there.

\vspace{1cm}
{\tt Acknowledgements}

\noindent 
This investigation was supported in part by the 
Hungarian National Science Fund OTKA (under T030099, T034299
and T043159).

\vspace{5cm}

\appendix
\section{Appendix}

In this appendix we recall the solution \cite{JKS,Kuniba} 
of the basic functional equation
\be
f(\xi+i)f(\xi-i)=B(\xi).
\label{fr}
\end{equation}
Here $f(\xi)$ is assumed to be meromorphic in the strip
$\vert{\rm Im\,}\xi\vert<1+\epsilon$, satisfying the reality 
condition
\be
f^*(\xi)=f(\xi^*).
\end{equation}
Let $\{r_\alpha\}_{\alpha=1}^R$ be the set of zeroes of $f(\xi)$
with $\vert{\rm Im\,}r_\alpha\vert<1$ (not necessarily different
from each other) and similarly $\{p_\beta\}_{\beta=1}^P$
the set of (not necessarily different) poles with 
$\vert{\rm Im\,}p_\beta\vert<1$. Moreover, for large (real, positive) $\xi$
\be
f(\xi)\sim\hat\sigma f_0\,\xi^{c_1}\,e^{c_0\xi}\,e^{-\lambda\cosh
\frac{\pi}{2}\xi}
\label{asf}
\end{equation}
asymptotically. Here $\hat\sigma=\pm1$ is a sign, $f_0>0$,  
$c_0,c_1$ and $\lambda$ are real.

The function $B(\xi)$ is real analytic in the strip
$\vert{\rm Im\,}\xi\vert<\epsilon$ and non-negative for real $\xi$.
Its asymptotics is given by
\be
B(\xi)\sim B_0\,\xi^{b_1}\,e^{b_0\xi},
\label{asB}
\end{equation}
where $B_0>0$ and $b_0,b_1$ are real.
Comparing (\ref{asf}) with (\ref{asB}) gives
\be
f_0=\sqrt{B_0},\qquad\qquad
c_1=\frac{b_1}{2},\qquad\qquad
c_0=\frac{b_0}{2}.
\end{equation}

The key observation that allows the transformation of the Y-system 
functional relations,
which are of the form (\ref{fr}), to TBA integral equations 
is the fact that the solution of (\ref{fr}),
satisfying all the above requirements is uniquely given by 
\be
f(\xi)=\hat\sigma\,e^{-\lambda\cosh\frac{\pi}{2}\xi}\,\,
\frac
{\prod_{\alpha=1}^R\tau(\xi-r_\alpha)}
{\prod_{\beta=1}^P\tau(\xi-p_\beta)}\,\,
\exp\{\beta(\xi)\},
\label{solFR}
\end{equation}
where
\be
\tau(\xi)=\tanh\left(\frac{\pi}{4}\xi\right)
\label{tau}
\end{equation}
and
\be
\beta(\xi)=\frac{1}{4}\,\int_{-\infty}^\infty
du\,\frac{\ln B(u)}{\cosh\frac{\pi}{2}(\xi-u)}.
\end{equation}
$\beta(\xi)$ is clearly analytic in the strip
$\vert{\rm Im\,}\xi\vert<1$ and for real $u$ (when $B(u)>0$)
\be
\beta(u\pm i)=\frac{1}{2}\ln B(u)\pm
\frac{i}{4}\,{\cal P}\,\int_{-\infty}^\infty
dv\,\frac{\ln B(v)}{\sinh\frac{\pi}{2}(v-u)}.
\end{equation}
Here ${\cal P}$ indicates principal value integration.


\begin{thebibliography}{99}
%
\bibitem{Luscher}
M. L\"uscher, Comm. Math. Phys. {\bf 104} (1986) 177;
{\bf 105} (1986) 153.
%
\bibitem{KM1}
T. R. Klassen and E. Melzer, Nucl. Phys. {\bf B362} (1991) 329.
%
\bibitem{LWW}
M. L\"uscher, P. Weisz and U. Wolff, Nucl. Phys. {\bf B359} (1991) 221.
%
\bibitem{YY}
C. N. Yang and C. P. Yang, J. Math. Phys. {\bf 10} (1969) 1115.
%
\bibitem{TS}
M. Takahashi and M. Suzuki, Prog. Theor. Phys. {\bf 48} (1972) 2187.
%
\bibitem{Zamo90}\
Al. B. Zamolodchikov, Nucl. Phys. {\bf B342} (1990) 695.
%
\bibitem{TBAlist}
T. R. Klassen and E. Melzer, Nucl. Phys. {\bf B338} (1990) 485;
{\bf B350} (1991) 635;\\
Al. B. Zamolodchikov, Nucl. Phys. {\bf B358} (1991) 497;
{\bf B366} (1991) 122;\\
V. A. Fateev and Al. B. Zamolodchikov, Phys. Lett. {\bf B271}
(1991) 91; \\
F. Ravanini, R. Tateo and A. Valleriani, Int. J. Mod. Phys. {\bf A8}
(1993) 1707;\\
R. Tateo, Phys. Lett. {\bf B355} (1995) 157;\\
P. Fendley, Phys. Rev. Lett. {\bf 83} (1999) 4468.
%
\bibitem{KP}
A. Kl\"umper and P. A. Pearce, J. Stat. Phys. {\bf 64} (1991) 13;
Physica  {\bf A183} (1992) 304;\\
A. Kl\"umper, M. T. Batchelor and P. A. Pearce, J. Phys. {\bf A24} (1991) 3111.
%
\bibitem{PCA}
P. A. Pearce, L. Chim and C. Ahn, Nucl. Phys. {\bf B601} (2001) 539;\\
{\tt hep-th/0302093}.
%
\bibitem{Fendley}
P. Fendley, Nucl. Phys. {\bf B374} (1992) 667.
%
\bibitem{Martins}
M. J. Martins, Phys. Rev. Lett. {\bf 67} (1991) 419.
%
\bibitem{DT}
P. Dorey and R. Tateo, Nucl. Phys. {\bf B482} (1996) 639;
{\bf B515} (1998) 575.
%
\bibitem{Fendley2}
P. Fendley, Adv. Theor. Math. Phys. {\bf 1} (1998) 210.
%
\bibitem{BLZ}
V. V. Bazhanov, S. L. Lukyanov and A. B. Zamolodchikov,
Comm. Math. Phys. {\bf 177} (1996) 381;
{\bf 190} (1997) 247;
{\bf 200} (1999) 297;
Nucl. Phys. {\bf B489} (1997) 487.
%
\bibitem{BE}
R. M. Ellem and V. V. Bazhanov, Nucl. Phys. {\bf B647} (2002) 404.
%
\bibitem{DdV0}
C. Destri and H. J. de Vega, Phys. Rev. Lett. {\bf 69} (1992) 2313;
Nucl. Phys. {\bf B438} (1995) 413.
%
\bibitem{DdV2}
C. Destri and H. J. de Vega, Nucl. Phys. {\bf B504} (1997) 621;\\
G. Feverati, F. Ravanini and G. Tak\'acs, Phys. Lett. {\bf B430}
(1998) 264; Nucl. Phys. {\bf B540} (1999) 543.
%
\bibitem{DdV1}
G. Feverati, F. Ravanini and G. Tak\'acs, Phys. Lett. {\bf B444}
(1998) 442.
%
\bibitem{DdVres}
D. Fioravanti, A. Mariottini, E. Quattrini and F. Ravanini,
Phys. Lett. {\bf B390} (1997) 243;\\
G. Feverati, F. Ravanini and G. Tak\'acs, Nucl. Phys. {\bf B570}
(2000) 615.

%
\bibitem{ZJ}
P. Zinn-Justin, J. Phys. {\bf A31} (1998) 6747.
%
\bibitem{JKS}
G. J\"uttner, A. Kl\"umper and J. Suzuki, Nucl. Phys. {\bf B512}
(1998) 581.
%
\bibitem{Kuniba}
A. Kuniba, K. Sakai and J. Suzuki, Nucl. Phys. {\bf B525 [FS]} (1998) 597.
%
\bibitem{LC}
C. Destri and H. J. de Vega, Nucl. Phys. {\bf B290} (1987) 363;
J. Phys. {\bf A22} (1989) 1329.
%
\bibitem{DTBA}
Al. B. Zamolodchikov, Phys. Lett. {\bf B253} (1991) 391;\\
E. Quattrini, F. Ravanini and R. Tateo, NATO ASI Series B:
Physics {\bf 328} (1995) 273.








\end{thebibliography}
\end{document}